\documentclass[aps,a4paper,superscriptaddress,twocolumn,pra]{revtex4-1}
\usepackage{graphicx}
\usepackage{amssymb}
\usepackage{amsfonts}
\usepackage{hyperref}
\usepackage{eucal}

\hypersetup{bookmarks=true}

\begin{document}

\title{Photonic chip based optical frequency comb using soliton induced Cherenkov radiation}

\author{V. Brasch}
\affiliation{ \'Ecole Polytechnique F\'ed\'erale de Lausanne (EPFL), CH--1015, Switzerland}

\author{T. Herr}
\affiliation{ \'Ecole Polytechnique F\'ed\'erale de Lausanne (EPFL), CH--1015, Switzerland}
\affiliation{Current address: Centre Suisse d'Electronique et Microtechnique SA (CSEM), CH--2002, Switzerland}

\author{M. Geiselmann}
\affiliation{ \'Ecole Polytechnique F\'ed\'erale de Lausanne (EPFL), CH--1015, Switzerland}

\author{G. Lihachev}
\affiliation{Faculty of Physics, M.V. Lomonosov Moscow State University, Moscow 119991, Russia}
\affiliation{Russian Quantum Center, Skolkovo 143025, Russia} 

\author{M.H.P. Pfeiffer}
\affiliation{ \'Ecole Polytechnique F\'ed\'erale de Lausanne (EPFL), CH--1015, Switzerland}

\author{M.L. Gorodetsky}
\affiliation{Faculty of Physics, M.V. Lomonosov Moscow State University, Moscow 119991, Russia}
\affiliation{Russian Quantum Center, Skolkovo 143025, Russia} 

\author{T.J. Kippenberg}
\email[Corresponding author: ]{tobias.kippenberg@epfl.ch}
\affiliation{ \'Ecole Polytechnique F\'ed\'erale de Lausanne (EPFL), CH--1015, Switzerland}

\begin{abstract}
Optical frequency combs\cite{Udem2002,Ye2005} provide a series of equidistant laser lines and have revolutionized the field of frequency metrology within the last decade. Originally developed to achieve absolute optical frequency measurements, optical frequency combs have enabled advances in other areas\cite{Newbury2011} such as molecular fingerprinting\cite{Diddams2007,Gohle2007b}, astronomy\cite{Steinmetz2008}, range finding\cite{Coddington2009} or the synthesis of low noise microwave signals\cite{Fortier2011}. Discovered in 2007\cite{Del'Haye2007, Kippenberg2011}, microresonator (Kerr) frequency combs have emerged as an alternative and widely investigated method to synthesize optical frequency combs offering compact form factor, chipscale integration, multi-gigahertz repetition rates, broad spectral bandwidth and high power per frequency comb line. Since their discovery there has been substantial progress in fundamental understanding\cite{Herr2012, Ferdous2011, Papp2011}, theoretical modeling\cite{Chembo2010,Matsko2011,Coen2013}, on-chip planar integration\cite{Levy2010,Moss2013} and resulting applications\cite{Pfeifle2014, Foster2011, Papp2014}. Yet, in no demonstration could two key properties of optical frequency combs, broad spectral bandwidth and coherence, be achieved simultaneously. Here we overcome this challenge by accessing, for the first time, soliton induced Cherenkov radiation\cite{Akhmediev1995, Wai1986} in an optical microresonator. By continuous wave pumping of a  dispersion engineered, planar silicon nitride microresonator\cite{Levy2010,Moss2013}, continuously circulating, sub-30\,fs short temporal dissipative Kerr solitons\cite{Akhmediev2008,Leo2010,Herr2013} are generated, that correspond to pulses of 6 optical cycles and constitute a coherent optical frequency comb in the spectral domain. Emission of soliton induced Cherenkov radiation caused by higher order dispersion broadens the spectral bandwidth to 2/3 of an octave, sufficient for self referencing\cite{Udem2002,Ye2005}, in excellent agreement with recent theoretical predictions\cite{Coen2013} and the broadest coherent microresonator frequency comb generated to date. Once generated it is shown that the soliton induced Cherenkov radiation based frequency comb can be fully phase stabilized. The overall relative accuracy of the generated comb with respect to a reference fiber laser frequency comb is measured to be  3 $\cdot$ 10 $^{-15}$.

The ability to preserve coherence over a broad spectral bandwidth using soliton induced Cherenkov radiation marks a critical milestone in the development of planar optical frequency combs, enabling on one hand application in e.g. coherent communications\cite{Pfeifle2014}, broadband dual comb spectroscopy\cite{Keilmann2004} and Raman spectral imaging\cite{Ideguchi2013}, while on the other hand significantly relaxing dispersion requirements for broadband microresonator frequency combs\cite{Lamont2013} and providing a path for their generation in the visible and UV.
Our results underscore the utility and effectiveness of planar microresonator frequency comb technology, that offers the potential to make frequency metrology accessible beyond specialized laboratories. 
\end{abstract}

\maketitle

Optical solitons are propagating pulses of light that retain their shape due to a balance of nonlinearity and dispersion\cite{Akhmediev2008,Leo2010,Grelu2012,Herr2013}. In the presence of higher order dispersion optical solitons can emit soliton Cherenkov radiation\cite{Wai1986,Akhmediev1995}. This process, also known as dispersive wave generation, is one of the key nonlinear frequency conversion mechanisms of coherent supercontinuum generation\cite{Dudley2006, Skryabin2010}, which allows substantially increasing the spectral bandwidth of pulsed laser sources.  The generation of a coherent supercontinuum from a pulsed laser propagating through an optical photonic crystal fiber, has been a defining experimental advance, which has enabled the first self-referenced optical frequency comb\cite{Udem2002,Ye2005} and has given access to coherent broadband spectra in the visible. This breakthrough has established the frequency comb technology with repetition rates up to around 1\,GHz now at the heart of a variety of applications\cite{Newbury2011,Ye2005}. 
However, a growing number of applications require pulse repetition rates in excess of 10\,GHz\cite{Newbury2011}. Beside the difficulty to generate multi-gigahertz repetition rates with conventional lasers, spectral broadening of such pulses is impeded by the inherent reduction of pulse energy with increasing repetition rate, which makes soliton spectral broadening inefficient.

One new route to these high repetition rate and broadband frequency combs was established with the discovery of microresonator frequency combs\cite{Del'Haye2007, Kippenberg2011}. Since the first experiments, the field of microresonator frequency combs, also known as Kerr frequency combs, has made substantial advances, such as the demonstration of frequency comb generation in CMOS compatible photonic chips using silicon nitride resonators\cite{Levy2010, Moss2013} as well as in other integrated microresonators\cite{Jung2013,Moss2013,Hausmann2014}, the achievement of electronically detectable microwave repetition rates\cite{DelHaye2008,Li2012}, the generation of octave spanning (yet high noise) spectra\cite{Del'Haye2011,Okawachi2011}, the extension to mid IR\cite{Wang2013,Griffith2015} {, generation of combs with pulsed output in normal dispersion resonators}\cite{Huang2015}{ and the ability to detect the carrier envelope offset frequency}\cite{Jost2014}. Proof of concept applications have been realized in coherent telecommunication\cite{Pfeifle2014}, miniature optical atomic clocks\cite{Papp2014}, optical waveform synthesis\cite{Ferdous2011} and low noise microwave generation\cite{Savchenkov2008}.  

Moreover a detailed understanding of the Kerr comb formation process has been obtained:
The Kerr comb formation process\cite{Herr2012} can lead  to multiple sub-comb formation in microresonators for which the variation of the free spectral range is small compared to the cavity linewidth; a parameter regime which in particular photonic chip based microresonators fall into. The formation of sub-combs can give rise to loss of coherence, multiple beat-notes and amplitude and phase noise.  
Low noise, coherent operation in this regime can still be achieved via sub-comb synchronization\cite{Herr2012}, which involves either probabilistic tuning techniques such as $\delta$-$\Delta$ matching\cite{Herr2012}, injection-locking\cite{DelHaye2014} and mode-locking\cite{Saha2013} or parametric seeding\cite{Papp2013}. 
In the opposite regime (which occurs in micro-toroid resonators\cite{Del'Haye2007} or crystalline resonators pumped in the mid IR\cite{Wang2013}) low phase noise can be achieved at reasonable pump power for narrow bandwidth frequency combs\cite{Ferdous2011,Papp2011,Herr2012}. Most recently temporal dissipative cavity (or dissipative Kerr) soliton formation has been observed in crystalline microresonators, leading to narrow, coherent frequency combs\cite{Herr2013}.

Recent theoretical advances based on models using the Lugiato-Lefever equation\mbox{\cite{Chembo2013, Coen2013,Lugiato1987}} have predicted that accessing solitons and soliton induced Cherenkov radiation\mbox{\cite{Coen2013,Erkintalo2014,Lamont2013}} provides a path to the reliable generation of broadband and coherent frequency combs, which can span a full octave, sufficient for self referencing. This regime therefore makes external broadening superfluous. 

\begin{figure}
\pdfbookmark[1]{Figure 1}{Fig.1}
\centering
	\includegraphics[width=0.45\textwidth]{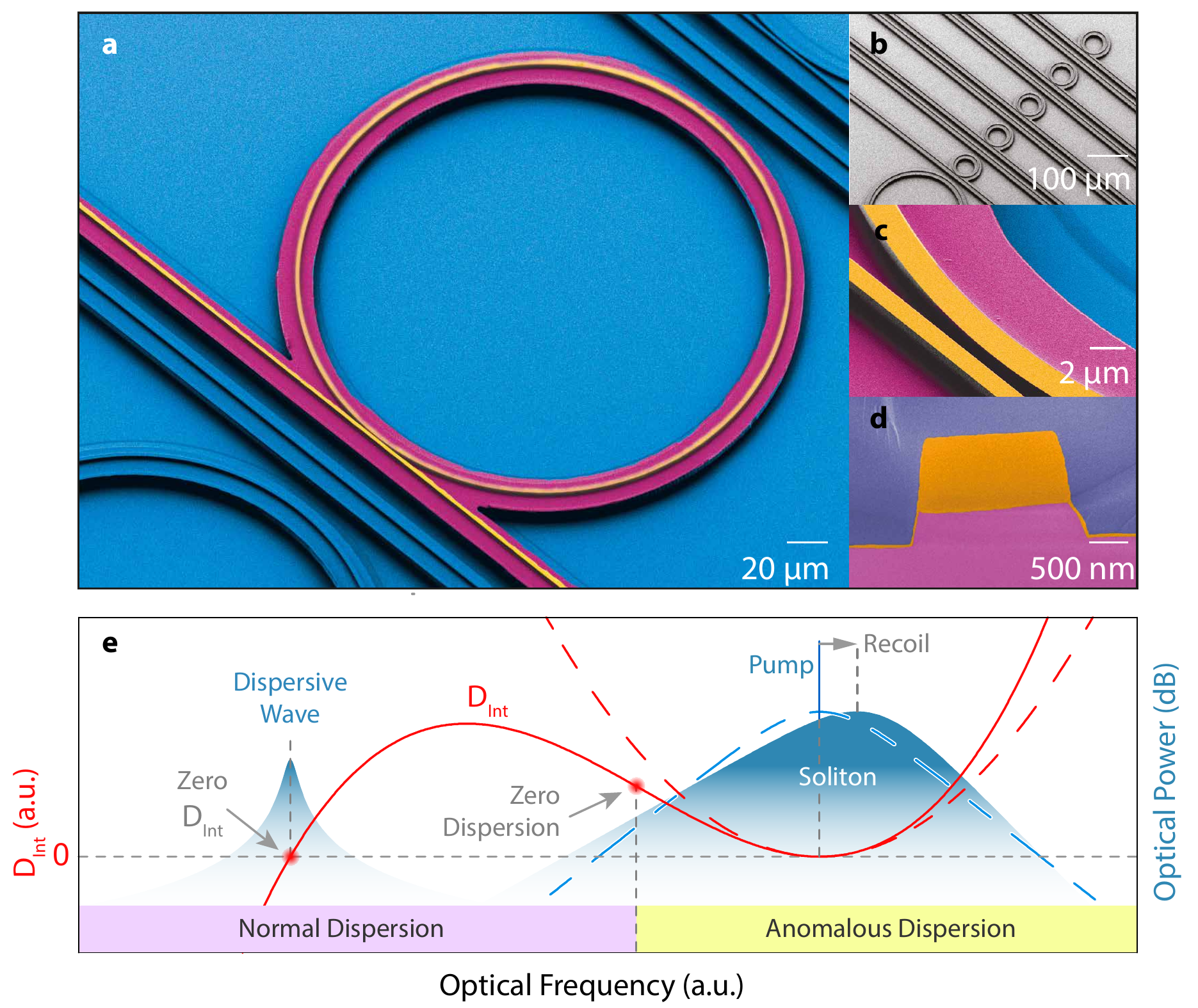} 
\caption{
\textbf{Temporal soliton generation and soliton induced Cherenkov radiation in a planar SiN microresonator. a,b} Colored scanning electron microscopy images of a SiN optical microresonator with the same geometry as the one used but without the SiO$_2$ encapsulation. Shown in blue is the silicon substrate. Colored in magenta is the SiO$_2$ pedestal with the SiN waveguide (orange) on top. \textbf{b,} An image of the chip at lower magnification. \textbf{c,} A close-up of the coupling region between bus waveguide and resonator (similar geometry as used). \textbf{d,} A cross section of a fabricated device that also shows the top cladding (SiO$_{2}$, colored purple). 
\textbf{e,} A schematic of the integrated dispersion $D_\mathrm{int}(\mu)$ as a function of optical frequency and the associated soliton dynamics.
Regions with positive (negative) curvature have anomalous (normal) group velocity dispersion (GVD). Around the pump $D_\mathrm{int}(\mu)$ can be approximated by a parabola (red dashed line), indicating that it is dominated by quadratic, anomalous GVD. Therefore the resonator supports the formation of temporal solitons around this wavelength. Due to the presence of higher order material and waveguide dispersion the region of anomalous GVD is finite. In the absence of fourth and higher order dispersion ($D_1, D_2, D_3 \neq 0$ and $D_p = 0$ for $p>3$) the zero dispersion point (ZDP) occurs at $\mu_\mathrm{ZDP} = - \frac{D_2}{D_3}$ (given by $D''_\mathrm{int}(\mu_\mathrm{ZDP})=0$).
Additionally, the integrated dispersion has at least one zero crossing, occurring at $\mu = -3\frac{D_2}{D_3}$, (given by $D_\mathrm{int}(\mu)=0$). In this case the soliton can emit soliton Cherenkov radiation (also known as dispersive wave emission) into the normal dispersion regime at around the frequency where $D_\mathrm{int}(\mu) = 0$ where phase matching occurs and the transfer of energy from the soliton to the dispersive wave is most efficient. To obey energy conservation in the presence of one strong dispersive wave the maximum of the generated spectrum close to the pump is shifted away from the pump (recoil).
}
\end{figure}

Our experimental platform are silicon nitride (Si$_3$N$_4$ further called SiN) optical microresonators, which are highly suitable for integrated nonlinear photonics due to its large band gap and the resulting wide transparency window\cite{Moss2013} and which are compatible with space applications\cite{Brasch2014}. Pioneering work\cite{Levy2010} demonstrated optical frequency comb generation in a SiN microresonator, providing a path to planar, CMOS compatible frequency comb sources.
Here we utilize 800\,nm thick SiN ring resonators with 238\,$\mu$m diameter embedded in SiO$_2$ (see Fig.\,1a-d and cf.\ Methods section) resulting in anomalous group velocity dispersion (GVD) around 1.5\,$\mu$m. The microresonator fabrication was optimized in order to mitigate avoided crossings of different mode families that can locally alter dispersion\cite{Herr2014}. 
The resonance frequencies of one mode family can be approximated around the pump resonance with eigenfrequency $\omega_0$ as a Taylor series $\omega_\mu = \omega_0 + \sum_j D_j \mu^j/j!$ with $j \in \mathbb{N}$, where $\mu \in \mathbb{Z}$ is the mode number counted from the designated pump resonance. Here $D_1/2\pi$ is the free spectral range of the resonator and $D_2$ is related to the GVD parameter $\beta_2$ by $D_2 = -\frac{c}{n}D_1^2\beta_2$. Figure 1e shows the integrated dispersion $D_\mathrm{int}(\mu)$ of a given mode number $\mu$ relative to the pump mode at $\mu=0$;  i.e. $D_\mathrm{int}(\mu) \equiv \omega_\mu-(\omega_{0}+D_1 \mu)=D_2 \frac{\mu^2}{2!} + D_3 \frac{\mu^3}{3!} +...$ . Therefore $D_\mathrm{int}(\mu)$ is equivalent to the deviation of $\omega_\mu$ from an equidistant frequency grid defined by $\omega_0$ and $D_1$.

When CW pumping an optical microresonator with anomalous GVD, bright temporal dissipative Kerr solitons can be formed\cite{Akhmediev2008, Leo2010, Herr2013}. 
The dynamics of this system can be described by a master equation\cite{Matsko2011a, Chembo2010, Coen2013} with higher order dispersion and self-steepening effects additionally taken into account:
\begin{widetext}
\begin{equation}
\frac{\partial A}{\partial  t}+i\sum_{j=2}\frac{D_j}{j!} \left(\frac{\partial}{i\,\partial \phi}\right)^jA - i g\left(1+\frac{iD_1}{\omega_0}\frac{\partial}{\partial \phi}\right) |A|^2A = \\
= -\left(\frac{\kappa}{2} +i(\omega_0-\omega_p)\right)A+\sqrt{\frac{\kappa\eta P_\mathrm{in}}{\hbar\omega_0}}.
\label{nls1}
\end{equation}
\end{widetext}

Here $A(\phi,t)=\sum_\mu A_\mu e^{i\mu\phi-i(\omega_\mu-\omega_p) t}$ is the slowly varying field amplitude, $\phi$ is the azimuthal angular coordinate inside the resonator, co-rotating with a soliton, $g = \frac{\hbar \omega_0^2 c n_2}{n^2 V_\mathrm{eff}} $ the nonlinear coupling coefficient with $n$ and $n_2$ the linear and nonlinear refractive indices of the material, $V_\mathrm{eff} = A_\mathrm{eff}  L$ the effective nonlinear mode volume with $A_\mathrm{eff}$ the effective nonlinear mode area and $L$ the cavity length, $\kappa$ the cavity decay rate, $\eta$ the coupling efficiency and $P_\mathrm{in}$ and $\omega_p$ the pump power inside the bus waveguide and the frequency of the pump light\cite{Herr2013}. 
Formally, equation\,\ref{nls1} is identical to the Lugiato-Lefever equation (LLE)\cite{Akhmediev2008} extended with higher order terms relevant for very short pulses\cite{Lamont2013}. In the frequency domain this equation is equivalent to a set of coupled mode equations, which directly describes the amplitudes of the modes in the resonator\cite{Herr2014} (cf. Methods).
 In the absence of third and higher order dispersion the LLE exhibits solutions that correspond to bright temporal solitons superimposed on a CW background\cite{Wabnitz1993}: 
 \begin{eqnarray}
  A(\phi) \approx A_\mathrm{cw} + A_1\sum^N_{j=1} \mathrm{sech}\left( \sqrt{\frac{2(\omega_0-\omega_p)}{D_2}} (\phi-\phi_j)\right)e^{i\psi_0}, 
 \end{eqnarray}
with amplitude $A_1$, phase $\psi_0$ and background $A_\mathrm{cw}$ determined by dispersion, nonlinearity, pump power and detuning, $\phi_i$ is the position of each soliton\cite{Herr2013}.
The minimal pulse duration at half maximum for a given pump power $P_\mathrm{in}$ is then given by 
$\Delta t_\mathrm{3dB} \approx 2 \sqrt{\frac{-\beta_2}{\gamma {\mathcal{F}} P_\mathrm{in}}}$, with ${\mathcal{F}}$  the finesse and $\gamma=\frac{\omega n_2}{c A_\mathrm{eff}}$ the effective nonlinearity.  These temporal dissipative Kerr solitons have been observed in fiber cavities\cite{Leo2010} and in crystalline microresonators\cite{Herr2013} and they occur only in the bistable regime, where simultaneous existence of the upper branch (soliton) and lower branch (CW pump) solution is warranted\cite{Herr2013}. 

When small higher order dispersion terms are present, the shape and velocity of the stationary solitons change\cite{Akhmediev1995}. The spectrum of such perturbed soliton becomes asymmetric with its maximum shifted from the pump frequency (cf. Fig.\,1e). At the same time the temporal soliton develops a radiation tail\cite{Akhmediev1995, Jang2014} which is the soliton Cherenkov radiation. {This is in contrast to dispersive wave formation in the regime of unstable modulation instability that can be explained by phase matching of cascaded four wave mixing\cite{Erkintalo2012}}. Due to the synchronization of the soliton and the Cherenkov radiation, this radiation can become comparable in strength to the soliton itself and can not be treated as a small perturbation\cite{Wai1986,Kuehl1990}. In this case another approach based on the linearization of the LLE may be used\cite{Skryabin2014} allowing to get complex valued $\mu_\mathrm{DW}$ for the position of the soliton Cherenkov radiation. Of the complex value the imaginary part is related to the width of the feature and the real part gives its position. This position is approximately given by the linear phase matching condition\cite{Erkintalo2012} that occurs for $\mu_\mathrm{DW} = -3\frac{D_2}{D_3}$ for $D_4 = 0$. In the presence of $D_4$ two peaks of Cherenkov radiation may occur near 
$ \mu_\mathrm{DW}= -\frac{2D_3}{D_4} \pm \sqrt{(\frac{2D_3}{D_4})^2- \frac{12D_2}{D_4} }  $. 

\begin{figure*}
\pdfbookmark[1]{Figure 2}{Fig.2}
\centering
\includegraphics[width=0.85\textwidth]{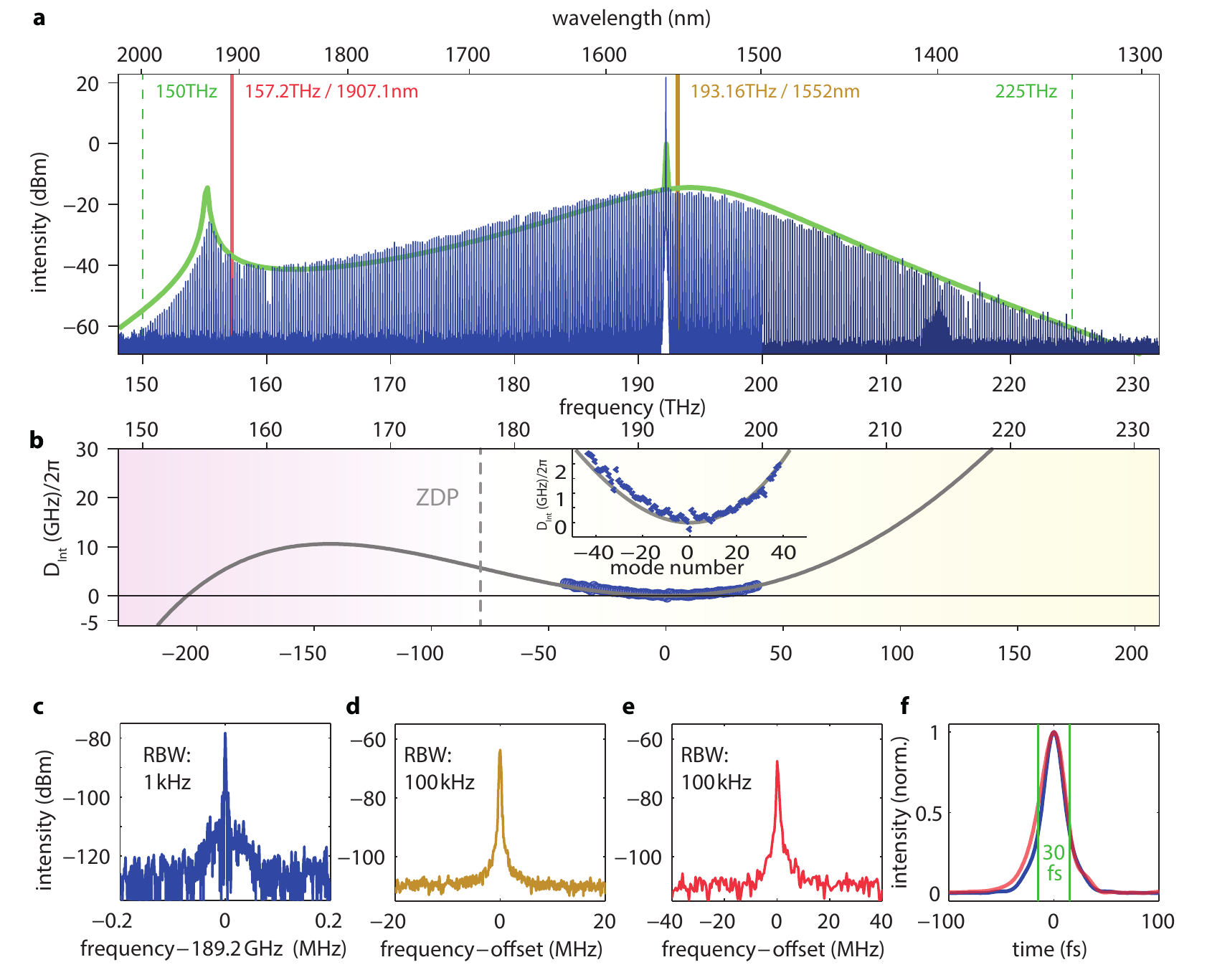}
\caption{\textbf{Single optical soliton and soliton Cherenkov radiation in a SiN optical microresonator} \textbf{a,} The optical spectrum shows the characteristic shape of a single soliton state in the microresonator with soliton Cherenkov radiation. The envelope of the spectrum is smooth and follows for the higher frequency side a sech$^2$-shape with a 3--dB width of 10.8\,THz. The Cherenkov radiation manifests itself as a sharp, distinct peak at around 155\,THz. The CW-pump is located at 192.2\,THz, the line spacing is 189.2\,GHz. The green dashed lines mark two frequencies between which the spectrum covers 2/3 of an octave. The green solid line represents the spectral envelope derived from simulations. {The two slightly different blues indicate measurements done with two different optical spectrum analyzers.} \textbf{b,} The integrated dispersion ($D_\mathrm{int}(\mu) =D_2 \frac{\mu^2}{2!} + D_3 \frac{\mu^3}{3!} +...$) from FEM simulations for the measured resonator geometry (grey solid line). The zero dispersion point (ZDP) is indicated by a dashed line. To the left of the ZPD is the normal group velocity dispersion (GVD) and to its right the anomalous GVD regime. The blue dots around mode number 0 (inset shows a zoom-in) are measured positions of around 80 resonances and show good agreement between the simulated dispersion and the measurements. \textbf{c,} The repetition rate beat note of the frequency comb at the line spacing of 189.22\,GHz shows a narrow linewidth of around 1\,kHz (cf. Fig.\,2b and Methods for details). \textbf{d,} The measured beat note of the generated frequency comb with a narrow linewidth reference laser positioned at 1552.0\,nm (orange line in \textbf{a}); offset is 1.92\,GHz. \textbf{e,} The beat note of a fiber laser positioned at 1907.1\,nm (red line in \textbf{a}) and the nearest comb line; offset is 5.47\,GHz. \textbf{f,} The intensity profile of the soliton pulse inside the resonator estimated from the measured spectrum (blue) and taken directly from the numerical simulation that yields the envelope in \textbf{a} (green) with FWHM of below 30\,fs. The pulse shape taken from the simulation shows a small asymmetry due to the effect of the Cherenkov radiation. }
\end{figure*}  

  In Fig.\,2b  frequency comb assisted diode laser spectroscopy\cite{Del'Haye2009} measurements of the SiN resonator dispersion are shown, revealing that around the pump wavelength the mode structure closely approaches a purely anomalous GVD (albeit very weak avoided crossings are still measurable, cf.\ inset and SI Fig.\,6, that are however not expected to impact soliton formation\cite{Herr2014}), fulfilling the requirements for bright soliton formation discussed above. The experimentally determined value $D_2/2\pi=$\,2.4\,$\pm$\,0.1\,MHz  is in close agreement with the value of $D_2/2\pi=$\,2.6\,MHz obtained by finite element modeling (FEM) based on the resonator geometry. 
When pumping the resonator at 1560\,nm via the bus waveguide with $P_\mathrm{in} \sim$\,500\,mW, far above the parametric threshold as given by the cavity bifurcation criterion (i.e. $\gamma \mathcal{F} P_\mathrm{thres} c/n \pi \approx \kappa/2$), we observe features that are canonical signatures of microresonator temporal soliton formation. While scanning the pump laser from higher frequencies to lower frequencies over a resonance of the TM$_{11}$ mode family a series of abrupt transmission steps occur. These discontinuities are also observed in the generated frequency comb light, (cf. SI Fig.\,9a and 5 for data from several other SiN microresonators).  Such  steps have been previously identified with temporal soliton formation in crystalline microresonators\cite{Herr2013}, where each step corresponds to the successive reduction of the number of solitons circulating in the microresonator. The regime where steps are recorded coincides with a narrowing of the recorded repetition rate beat note (cf.\ SI Fig.\,5c,d) in agreement with the low noise operation within the soliton formation regime.  These observations thus indicate temporal soliton formation in SiN microresonators.
 
To access the soliton states in a steady state, we developed a laser tuning technique (cf.\ Methods) to overcome instabilities associated with the discontinuous transitions of the soliton states. After tuning into the soliton state it remains stable for hours without any further stabilization allowing to record the full optical spectrum and to investigate its coherence properties. 
The optical single soliton spectrum for approximately {2}\,W of pump power in the bus waveguide is shown in Fig.\,2a and has several salient features. First, it covers a bandwidth of 2/3 of an octave, from 150\,THz to 225\,THz. Second, it exhibits the characteristic hyperbolic secant spectral envelope near the pump that is associated with a temporal soliton. The 3-dB bandwidth of the generated spectrum is 10.8\,THz which corresponds to $29$\,fs optical pulses (i.e. a pulse with 6 optical cycles). This agrees well with the expected bandwidth from the calculated minimum soliton duration of $\Delta t_\mathrm{3dB} \approx 25$\,fs. The soliton pulse energy is estimated to be 0.1\,nJ inside the resonator. Third, a striking attribute is the sharp feature around 1930\,nm (155\,THz) that corresponds to the soliton induced Cherenkov radiation\cite{Coen2013}. Figure 2b shows the measured dispersion and the dispersion simulated with the aforementioned finite element modeling (cf.\ Methods). The spectral position of the Cherenkov radiation at $\mu=195$ is in very good agreement with the linear phase matching condition that occurs at $\mu_\mathrm{DW}\approx 200$ for the simulated parameters  $D_2/2\pi=  2.6$\,MHz, $D_3/2\pi= 24.5$\,kHz, $D_4/2\pi= -290$\,Hz.  
  
We compared the measured single soliton spectrum to numerical simulations of the soliton dynamics\cite{Herr2013} using dispersion parameters derived from FEM simulations. The resulting spectrum is shown as the envelope in Fig.\,2a and again, good agreement is attained. From the simulations we can also calculate the temporal shape of the soliton inside the microresonator, which agrees well with the estimate based on the measured spectrum assuming flat phases (cf.\ Fig.\,2f). The good agreement with experimental data therefore establishes numerical simulations\cite{Coen2013,Lamont2013} as a powerful predictive tool for soliton dynamics in microresonators. Noticeable differences between the experimental spectrum and the theoretical simulation are the reduced intensity of the Cherenkov radiation and the absence of an observable soliton recoil in the experiment. We attribute the former to experimentally confirmed higher optical losses in the setup for these long wavelengths and the latter to possible variations in the quality factor versus wavelength and nonlinear terms ({Raman shift\cite{Milian2015}}, frequency dependence of the nonlinear coefficient), which are not taken into account in our simulation.

 A key property of a frequency comb, the coherence of its spectrum, has to be investigated separately. First, the repetition rate beat note of the generated soliton spectrum was measured on a high-speed photodiode at a frequency of 189.2\,GHz using amplitude modulation down-mixing\cite{DelHaye2012} (cf. Methods and SI Fig.\,9b).
 Figure 2c shows the resulting resolution bandwidth limited beat note (resolution bandwidth of 1\,kHz)  which exhibits a signal to noise ratio (SNR) of 40\,dB in 100\,kHz bandwidth. This demonstrates the low noise nature of the temporal soliton state. In addition, we record the low frequency amplitude noise of the soliton state and find no excess noise compared to the pump laser noise (cf.\ SI Fig.\,8). These measurements demonstrate that the soliton component of the spectrum centered around the pump exhibits low noise and is fully coherent.
 
Given that {in theory an interaction of the soliton and its radiation tail was studied\cite{Yulin2013}, potentially leading to loss of coherence,} it demands for a detailed analysis of the coherence of the soliton Cherenkov radiation in the microresonator case. To do so we carry out additional CW heterodyne beat note measurements with a thulium fiber laser (wavelength 1907\,nm). 
To reduce the effect of frequency jitter of the pump on the measurements, we stabilize the pump laser onto an absolute frequency reference.
 Figure 2e shows the beat of the thulium fiber laser at 1907\,nm with one frequency comb component of the soliton induced Cherenkov radiation, exhibiting a narrow linewidth $\sim$\,1\,MHz and $>20 $\,dB high SNR in 2\,MHz bandwidth.  Simultaneously with the beat at 1907\,nm we measure the beat with an erbium fiber laser close to the pump at 1552\,nm. The resulting CW beat is similar in width to the in-loop beat of the stabilized pump laser ($\sim$\,300\,kHz). {We attribute the difference between the widths of the beat at 1552\,nm and 1907\,nm to the effect of the non-stabilized repetition rate.}
The narrow beat note for the line spacing together with the two heterodyne measurements proof that the entire spectrum is coherent over its full span of 2/3 of an octave, in contrast to earlier reports\cite{DelHaye2008,Okawachi2011}. We note that this spectrum constitutes the broadest coherent spectrum generated {directly on chip} to date and that its bandwidth can be increased further by dispersion engineering\cite{Lamont2013}.
 
 It is insightful to contrast the coherent single soliton state to the incoherent high noise state that can be generated when tuning the pump laser continuously into the resonance, a tuning mechanism that has been widely employed in Kerr frequency comb generation experiments.  We observe in this case a spectrum that markedly deviates in its shape from the single soliton spectrum (see SI Fig.\,8a). While the overall bandwidth is only slightly reduced the spectrum is not coherent and a degradation of the beat note {to a width of the order of GHz} is measurable in CW heterodyne measurements {(SI Fig.\,8b)}, in agreement with the formation of sub-combs\cite{Herr2012}. In this high noise regime the height of the Cherenkov radiation is significantly lower (by approx. 10\,dB) and the spectrum develops two characteristic lobes symmetric around the pump. Also this low coherence state is in good agreement with our numerical simulations (cf. SI Fig.\,8a) and recent theoretical predictions\cite{Erkintalo2014}.
  
\begin{figure}
\pdfbookmark[1]{Figure 3}{Fig.3}
\centering
\includegraphics[width=0.5\textwidth]{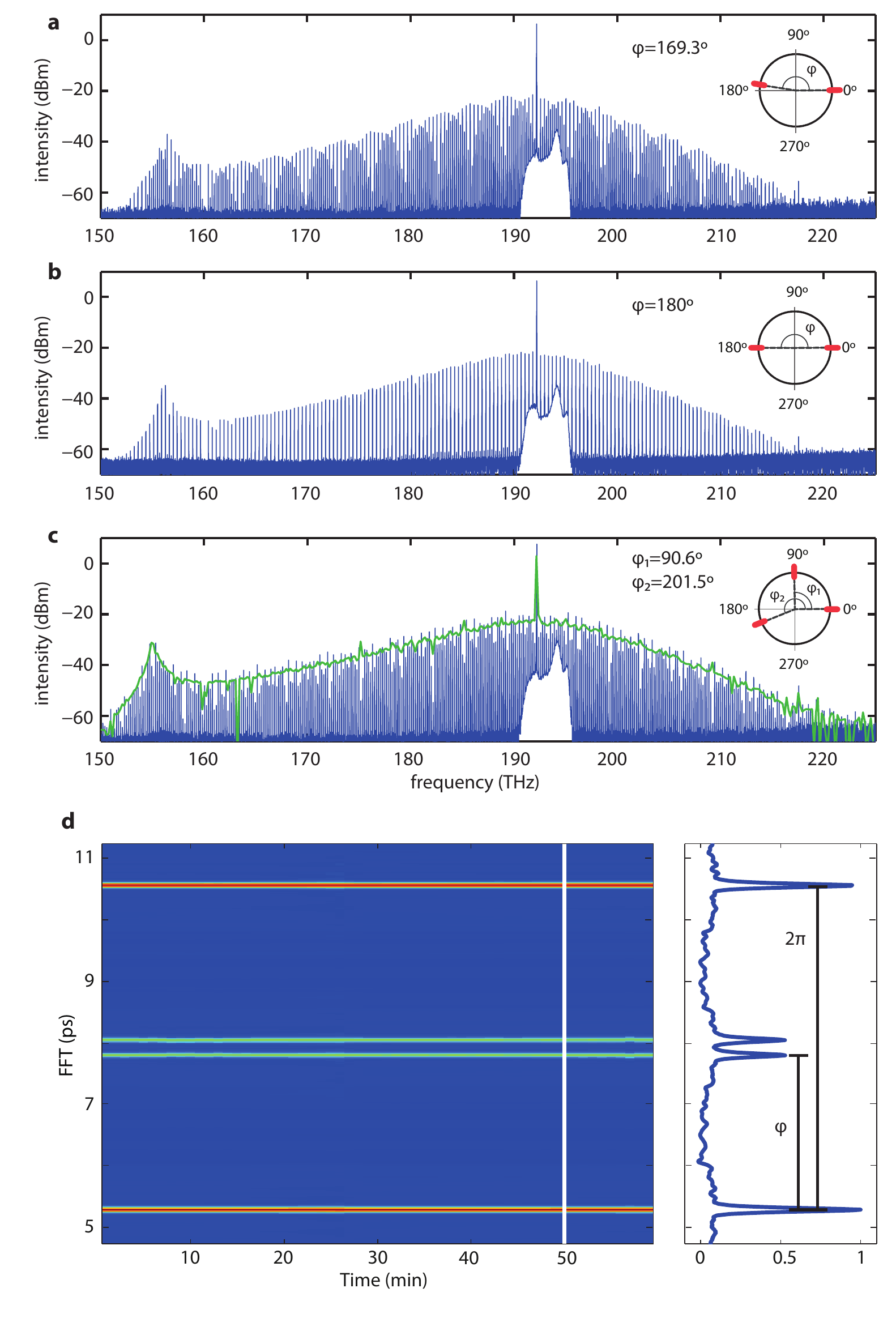}
\caption{
\textbf{Multi-soliton states in a planar microresonator and their stability.} \textbf{a,b,c} Spectra for multi-soliton states and the relative phase position of the solitons inside the microresonator, sketched in the insets according to the field autocorrelation (Fourier transform of the intensity spectrum). \textbf{a,b} Two soliton states and \textbf{c} three soliton state with the derived single soliton spectral envelope  (solid green line). \textbf{d,} The color-coded field autocorrelation of a two-soliton state versus laboratory time. The peaks (bright blue and red horizontal lines) give directly the positions of the two solitons with respect to each other. The position jitters only very little and does not drift over the time of one hour. The panel to the right shows the field autocorrelation after 50\,min indicated by the white line.}
\end{figure}  

Our system also allows to study multiple temporal solitons in the cavity which are stable for hours, {as the single soliton state}, (Fig.\,3d) allowing for the analysis of their coherence properties. Figure 3a--c show the optical spectra of three multi-soliton states, which we find again to be coherent (cf.\ SI Fig.\,7). The generated spectra show pronounced variations in the spectral envelope, that arise from the interference of the Fourier components of the individual solitons. The spectral envelope function of this interference is given by 
\begin{eqnarray*}
I(\mu)=\left|\sum\limits_{j=1}^{N} \mathrm{exp}(i\phi_j \mu)\right|^2,
\end{eqnarray*}
where $\phi_j$ corresponds to the relative angular position of the $j^\mathrm{th}$ soliton. 
The insets of Fig.\,3a--c show the reconstructed relative positions of the solitons inside the resonator for the different spectra (cf.\ Methods). Figure 3b shows the case where two solitons are almost perfectly opposite to each other in the resonator (differing by 180 degrees). This results in a spectrum that exhibits twice the line spacing of the single soliton case. Figure 3c shows that a higher number of cavity solitons ($N=3$) can result in a spectrum with more complex spectral modulations. Using the soliton positions, we can retrieve successfully the equivalent single soliton spectrum (solid green envelope in Fig.\,3c).

\begin{figure}
\pdfbookmark[1]{Figure 4}{Fig.4}
\centering
\includegraphics[width=0.45\textwidth]{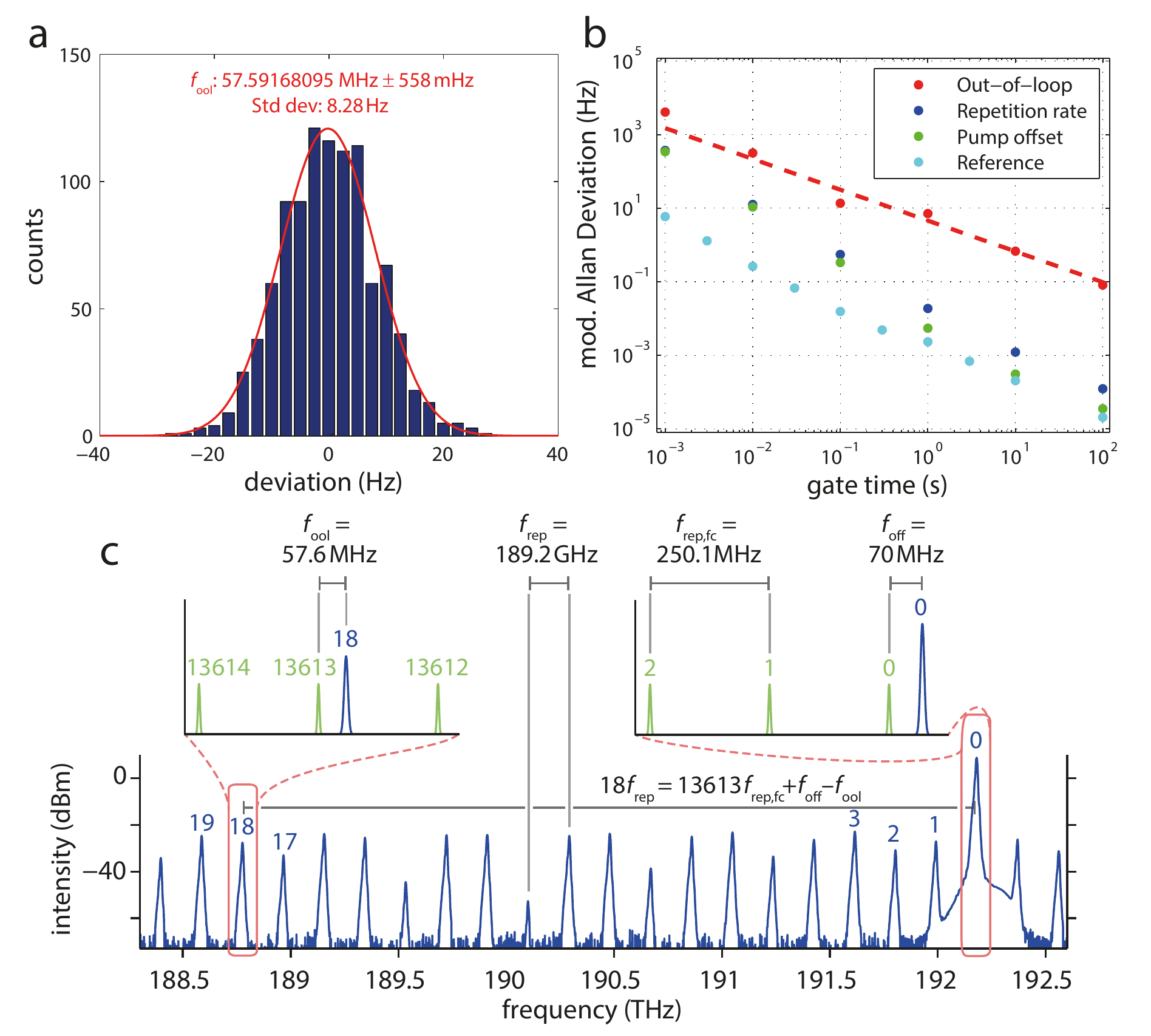}
 \caption{\textbf{Full phase stabilization and absolute frequency accuracy measurement of dissipative Kerr solitons in SiN. a} Histogram of the counter measurement for the out-of-loop beat of the stabilized microresonator frequency comb with a commercial fiber laser frequency comb. Gate time is 1\,s. The Gaussian fit gives the exact frequency of the beat ($f_{\text{ool}}$). The stabilized state shown here is a two soliton state. \textbf{b} The modified Allan deviation of the out-of-loop beat as well as the in-loop signals for the two locks of the repetition rate and the pump laser offset of the microresonator frequency comb. All signals average down over the gate time as expected for coherent signals. \textbf{c} A scheme highlighting the principle of the absolute frequency accuracy measurement referenced to a self-referenced fiber frequency comb. The out-of-loop beat is the beat between the 18$^{\text{th}}$ line on the red side of the pump of the microresonator frequency comb and the 13613$^{\text{th}}$ line of the reference comb counted from the line that the pump laser is locked to. The exact frequencies are given in the Methods section.}
\end{figure}

To prove the usability of our system for metrological applications we implement a full phase stabilization of the spectrum by phase locking the pump laser and the repetition rate of the SiN comb to a common RF reference. For the absolute frequency stabilization of the pump laser we use an offset lock to a self-referenced fiber laser frequency comb. The actuation for the repetition rate is achieved via the pump power\cite{DelHaye2008} (cf. Methods). In Fig.\,4b we show the modified Allan deviation of the in-loop signals, i.e. the repetition rate and offset lock respectively for a two soliton based Cherenkov radiation comb state. To verify the stabilization we record an out-of-loop signal, by counting the heterodyne beat of an independent comb teeth of the SiN comb with the reference comb. For all three signals the modified Allan deviation averages down with increasing gate time. The slope of the modified Allan deviation for the out-of-loop data in Fig.\,4b is $-0.83$. Similar results have been measured for a single soliton state as well (cf. SI Fig.\,10). These measurements demonstrate that the soliton induced Cherenkov radiation state is stable against pump and laser fluctuations and allows a full phase stabilization.

The out-of-loop measurement also allows to compare the absolute frequency accuracy of the soliton Cherenkov radiation based comb state with the fiber laser reference comb. To assess the absolute accuracy we compare the 18$^{\text{th}}$ soliton comb teeth to the fiber laser comb (with approximately 250\,MHz mode spacing). After numerically infering that the beat with the fiber comb originates from the 13613$^{\text{th}}$ comb teeth (relative to the pump) we can measure the absolute frequency difference $\Delta$ of the two intervals, and thus access the soliton comb accuracy. Taking into account all locked frequencies as shown in Fig.\,4c and extracting the center frequency of the out-of-loop signal from counter measurements shown in Fig.\,4a, we derive a frequency difference of $\Delta = 18 \cdot f_{\text{rep}} - 13613 \cdot f_{\text{rep,fc}} - f_{\text{off}} + f_{\text{ool}} = 25 \pm 558$\,mHz for the 1000\,s long measurement. We therefore validate the accuracy (and thereby also the equidistance) of the SiN soliton frequency comb to sub-Hz level and verify the relative accuracy (with respect to the optical carrier) to $3 \cdot 10^{-15}$. 

The observation of broadband and coherent spectra created by soliton formation and soliton induced Cherenkov radiation in a photonic chip based microresonator provides an essential novel ingredient to realize on chip frequency combs for two reasons. First, the mechanisms are well understood which makes quantitative predictions of the generated coherent spectra possible. Second, they expand the frequency comb bandwidth to the normal GVD window which can extend the frequency comb bandwidth to the visible wavelength range, a regime in which most materials have normal GVD.
The presently achieved coherent 2/3 of an octave with detectable mode spacing can be self-referenced with the 2\textit{f}-3\textit{f} technique\cite{Ye2005,Udem2002} without any additional broadening and can be extended to a full octave with a modified dispersion design, as previously predicted\cite{Lamont2013}.  
From a time domain perspective, the  approach enables the synthesis of few cycle light pulses, that are already at present shorter than conventional fiber laser technology based on erbium.
The generated broadband frequency comb spectra with multi-GHz mode spacing can be directly used for astrophysical spectrometer calibration, meeting the specifications for e.g. the SPIROU spectrometer, employed for the search of earth like exoplanets. 
In conjunction with the ability to integrate lasers and detectors on the same chip our results provide a key step towards the long term goal of a fully integrated, on-chip RF-to-optical link.

\begin{acknowledgements}
This work was supported by the DARPA Program QuASAR, an ESA TRP and the Swiss National Science Foundation. MLG acknowledges support from the RFBR grant 13-02-00271.
Correspondence and requests for materials should be addressed to TJK~(email: tobias.kippenberg@epfl.ch).
\end{acknowledgements}

\section*{Methods}
\subsection{Nanofabrication}
Starting with a silicon wafer with 4\,$\mu$m of thermal silicon dioxide (SiO$_2$) we deposit close to stoichiometric silicon nitride (SiN) for the waveguide cores as a 800\,nm thick film in a low-pressure chemical vapor deposition (LPCVD) process. After some auxiliary steps the waveguides are patterned in a 100\,kV electron beam lithography system using ZEP520 as resist. After development the resist is re-flown. The following reactive ion etch is the critical etch step. It uses CHF$_3$ and SF$_6$ gases and transfers the pattern into the SiN. This is followed by a photolithography step to define auxiliary structures. The wafer is thoroughly cleaned before an additional thin layer of SiN is deposited with the same process as before. Afterwards the wafer is annealed and the 3\,$\mu$m thick SiO$_2$ cladding is deposited with a CVD process on top. The last steps are the definition and separation of the chips and a second anneal.

\subsection{Finite element simulations}
Using the commercial Comsol Multiphysics package for FEM simulations, we implemented a 2D simulation which takes into account the cylindrical symmetry of the system in the third dimension. Material dispersion is taken into account via an iterative approach which takes the values of the refractive index from measured values for our SiN films. The values for dispersion parameters are obtained from fitting appropriate polynomials to the absolute frequencies of the modes around our pump wavelength of 192.2\,THz. The relative magnitude of the values for the free spectral range in the simulation is used to identify the mode families (TE and TM) inside the resonator.

\subsection{Implementation of the numerical simulation}
For the numerical simulation we used coupled mode equations which are propagated in time using an adaptive step size Runge-Kutta algorithm\cite{Chembo2010}. The calculation of the nonlinear mixing terms is efficiently calculated in the Fourier-domain\cite{Hansson2014}. An additional self-steepening term\cite{Lamont2013} was added to better model the behavior of few-cycles pulses. To allow for deterministic simulations of required states one, two or three-soliton states were seeded as initial waveform, starting the simulation with a detuning that allows for stable solitons.
The parameters used for the simulation presented in Fig.\,2a are:\ $D_2 /2\pi=$\,2.2\,MHz, $D_3/2\pi = $\,25\,kHz, $D_4/2\pi = -300$\,Hz, detuning $\zeta=$\,12, $Q_{\text{int}} =$\,1.5\,$\cdot$\, 10$^6$, $Q_{\text{ext}} =$\, 8\,$\cdot$\,10$^5$, $P_{\text{pump}} =$\,1\,W.

\subsection{Multi-soliton and single-soliton spectra}
When $N$ identical solitons circulate around the resonator they produce a frequency comb spectrum $S^{(N)}(\mu)$ with a structured envelope which results from interference of single soliton Fourier spectra $S^{(1)}(\mu)$: 
\begin{widetext}
\begin{equation}
S^{(N)}(\mu)=\left|{\cal F}\left\{\sum_{j=1}^N A_\mu(\phi-\phi_j)\right\}\right|^2= \\
S^{(1)}(\mu)I(\mu)=S^{(1)}(\mu)\left(N+2\sum_{j\neq l} \cos(\mu(\phi_j-\phi_l))\right). 
\end{equation}
\end{widetext}
If now the Fourier transform of this optical spectrum (sometimes transformed initially by picking only the frequency comb lines) is taken, it will result in the field autocorrelation function of the waveform according to the Wiener-Khinchin theorem. In the case of $N$ solitons it contains two peaks at $0$ and $2\pi$ and in addition $N(N-1)$ single-soliton autocorrelation peaks at positions in the interval (0..2$\pi$) symmetric around $\pi$, corresponding to pairwise distances between circulating solitons (cf.\ Fig.\,3d). Using simple peak-finding it is possible to obtain these distances and define the trigonometric multiplier of the spectrum. To reconstruct the single-soliton spectrum we simply divide the initial spectrum by
$N+2\sum_{j\neq l}^N \cos(\mu(\phi_j-\phi_l))$. Note that in the degenerate case when all the distances are related as integer numbers (like in Fig.\,3b) the trigonometric sum may turn to zero at some points, which should be dropped from consideration. With the measured or reconstructed single-soliton spectrum and assuming flat phases it is possible to estimate an approximate symmetrized form of the soliton and its duration, assuming that the asymmetry of the pulses is small (Fig.\,2f). From the 3--dB width of the spectrum the pulse duration can be derived via the time-bandwidth product for soliton pulses of 0.315\cite{Herr2013}.

\subsection{Sample characterization}
The dispersion of our resonators was measured by sweeping a widely tunable external cavity diode laser (ECDL) over the resonances and recording the transmitted power on a photodiode. Part of the power of the ECDL is differed before the resonator and used to record its beat signal with a commercial fiber laser frequency comb (repetition rate around 250\,MHz). By counting the crossings of this beat at certain frequencies and interpolation the position of the laser and the relative positions of the resonances can be determined with a precision of a few MHz\cite{Del'Haye2009}.
The average {linewidth of 300\,MHz} of the mode family used for the generation of the frequency comb is measured by determining the linewidths of many resonances within the range from 1510\,nm to 1580\,nm. To determine the linewidth the laser is scanned over each resonance and the polarization is optimized. The laser scan is calibrated by using the same technique that is used to calibrate the laser scan for the dispersion measurement above.
To measure the parametric threshold an amplified diode laser is swept over the resonance while the pump light is filtered out from the transmission using a tunable fiber Bragg grating. The remaining converted light which is the light at other frequencies than the pump is detected on a photodiode. The power of the pump is adjusted until a clear signature of converted light on the photodiode is observed indicating that parametric frequency conversion takes place. To take into account asymmetric input and output losses from the chip this measurement is repeated with changed direction on the chip. The measured threshold is $P_{\text{thres}} =$\,300 mW in the waveguide. The coupling loss per chip facet is approximately 3\,dB.

\subsection{Laser tuning procedure for soliton generation and beat note measurements}
 In order to achieve a stable soliton state we have to overcome the transient instability of the states within the steps. This is achieved by modulating the pump power with a simple two step protocol. The first step is to induce the soliton with a quick drop in power. The second step is to stabilize the soliton state by increasing the pump power.
The two step process to obtain stable soliton states is implemented using one acoustic optical modulator (AOM) and one Mach-Zehnder amplitude modulator (MZM). The only reason for the use of two modulators is the required speed of the modulation which can not be obtained solely with the AOM. The first step of a short dip in power is done with the MZM and typically of 100 to 200\,ns in length. The following increase in pump power is obtained with the AOM as a step function that remains at high power throughout the measurements afterwards. 
In order to measure the electronic beat note at 189\,GHz with a photodiode, we {suppress the pump by around 30\,dB using a fiber Bragg grating and }modulate sidebands of 40\,GHz onto the lines of the {remaining } frequency comb using a MZM. The corresponding modulation sidebands reduce the difference to 109\,GHz amenable to direct detection with a commercial telecommunication photodiode  with an optical power of around 1\,mW. The electrical signal of 109\,GHz as obtained from the photodiode is down mixed with a harmonic mixer on the sixths harmonic before being detected with an electrical spectrum analyzer at a frequency of around 1.7\,GHz.
The two narrow linewidth fiber lasers used for the heterodyne beat note measurements are based on NKT Koheras sources and have linewidths below 10\,kHz and a very high frequency stability.

\subsection{Full stabilization}
The repetition rate is detected as described above and downmixed to 70\,MHz. The 70\,MHz signal of the repetition rate and the pump laser offset at the same frequency are fed into two separate digital phase comparators with the same 70\,MHz reference signal on both. The phase comparators provide the error signal for two PID controllers which provide the feedback which is used to modulate the pump power via an AOM and the laser current for the repetition rate lock and the laser offset lock respectively. All involved RF equipment in the scheme is referenced to the atomic reference of the fiber frequency comb that provides the absolute optical reference for the pump laser offset lock. To measure the modified Allan deviation, three counters of the $\Lambda$-type were used to measure the frequencies of the laser offset, the repetition rate and the out-of-loop beat simultaneous. The result therefore does not agree perfectly with the modified Allan deviation but shows the same scaling for different noise sources\cite{Dawkins2007}.

The exact frequencies are for the repetition rate of the microresonator frequency comb (locked, $f_{\text{rep}}$): 189179.658\,MHz, for the repetition rate of the fiber laser frequency comb used as a reference (locked, $f_{\text{rep,fc}}$): 250.144820075\,MHz, for the offset of the pump laser from the reference frequency comb (locked, $f_{\text{off}}$): 70.000\,MHz and for the out-of-loop beat between the two frequency combs (unlocked, derived from fit in Fig.\,4(a), $f_{\text{ool}}$): 57.59168095\,MHz.

\bibliography{V58_Arxiv}

\newpage
\clearpage

\appendix
\onecolumngrid
\section*{Supplementary Information}

\begin{figure*}[h]
\pdfbookmark[1]{Figure 5}{Fig.5}
\centering
\includegraphics[width=0.6\textwidth]{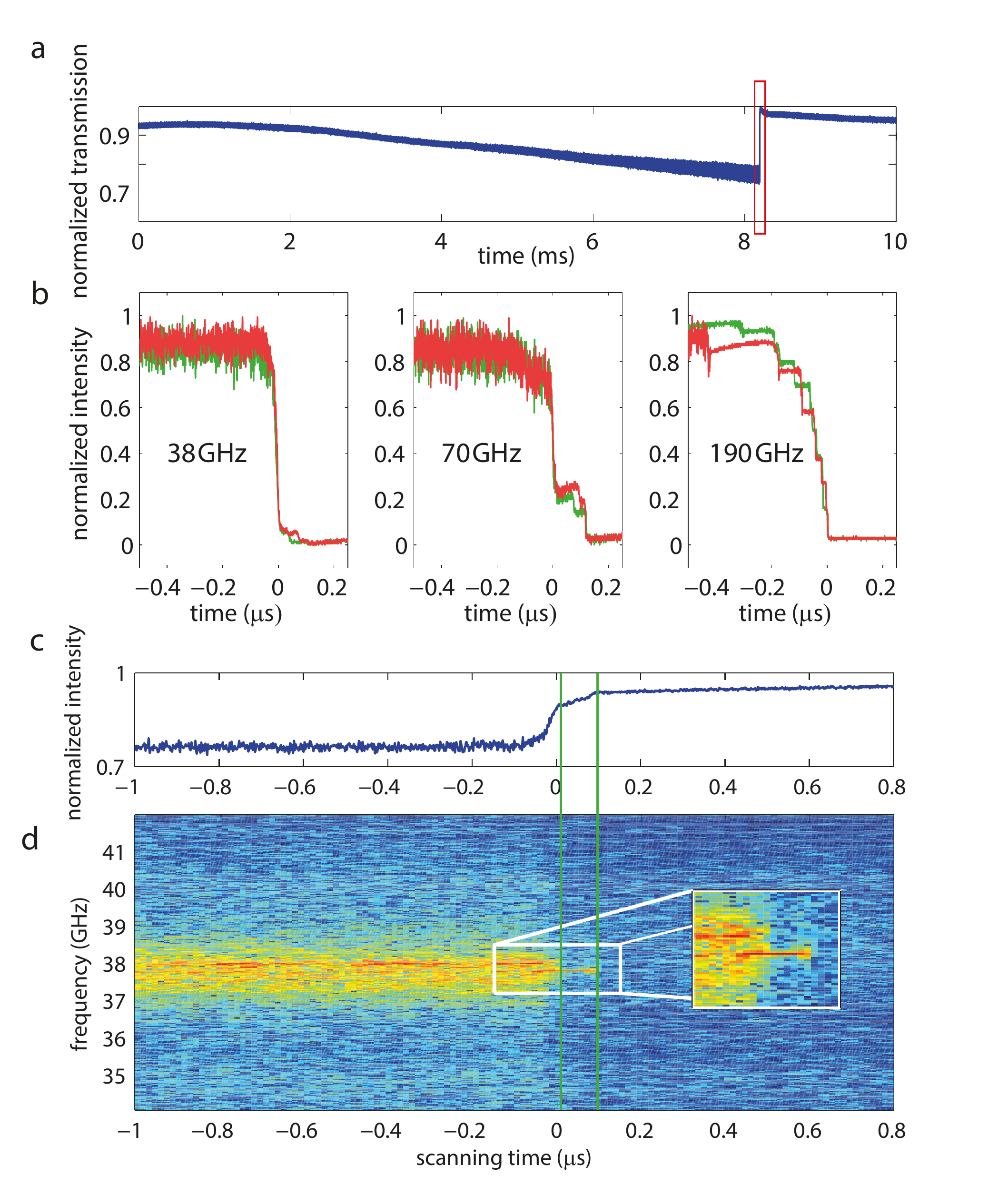}
\caption{\textbf{SI -- Signs of solitons in SiN microresonators.} \textbf{a,} The typical triangular shape of the transmission for higher power laser sweeps from the blue side to the red side of a resonance. \textbf{b,} The steps in the converted light at the end of the triangle (marked with a red box in \textbf{a}) are strong signs for soliton states. Red and green are two traces from two consecutive laser sweeps, highlighting the changes in the step patterns. Resonators with different repetition rates, as noted inside the figures, show these steps. \textbf{c,} The step in the transmission of a 38\,GHz sample with high time resolution at the position of the red box in \textbf{a}. \textbf{d,} The collapse of the beat note at 38\,GHz from around 1\,GHz width to a width limited by the measurement technique within the short step of \textbf{c}.}
\end{figure*}

\begin{figure*}[h]
\pdfbookmark[1]{Figure 6}{Fig.6}
\centering
\includegraphics[width=0.7\textwidth]{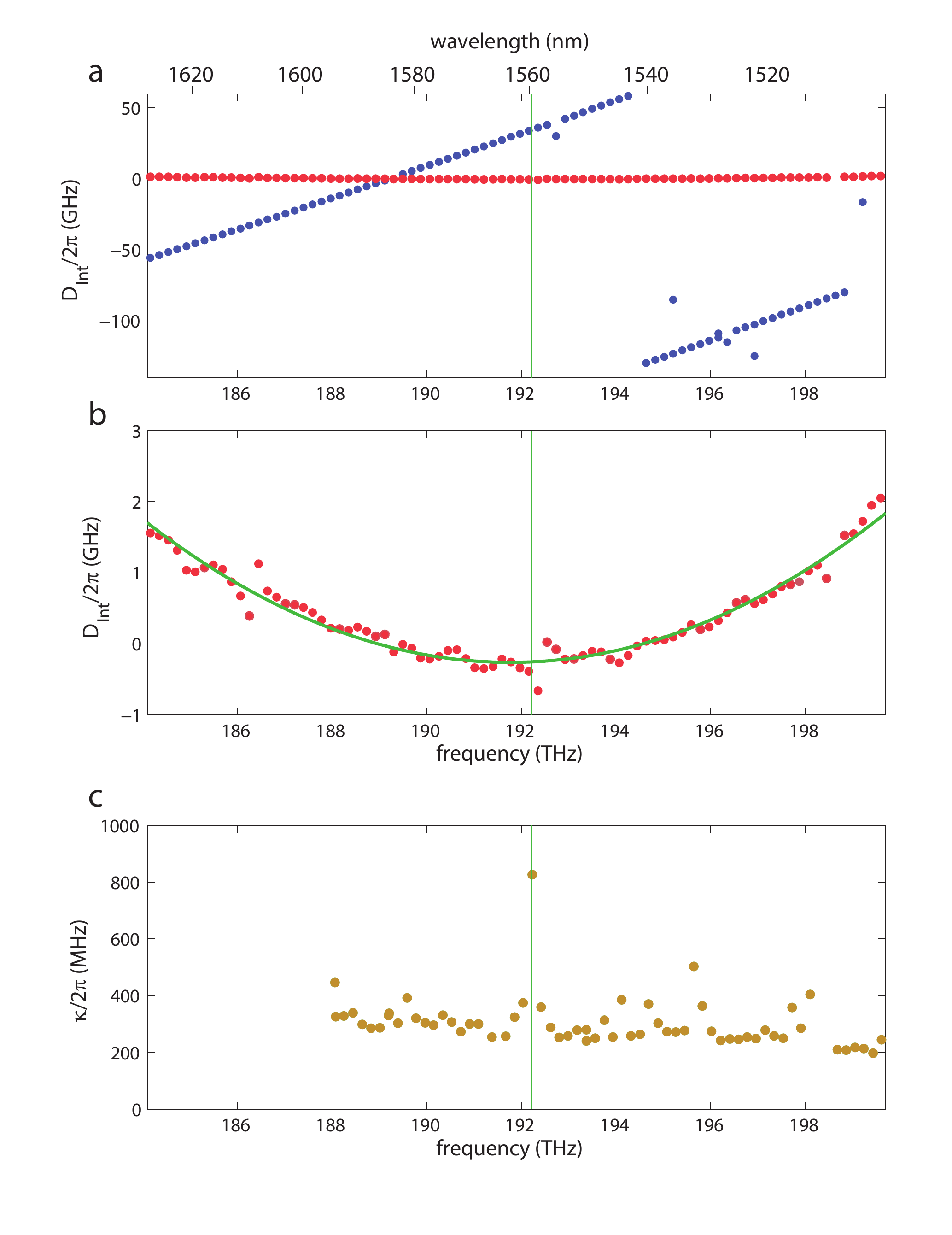}
\caption{\textbf{SI -- Sample characteristics.} \textbf{a,b} $D_\mathrm{int}$ of resonances in the used sample. The red, horizontal mode is the pumped TM$_{11}$ mode (enlarged in \textbf{b}). The blue TE$_{11}$ mode has a different FSR and therefore is a tilted line. The solid, green line in \textbf{b} is a quadratic fit yielding a value for D$_2/2\pi$ of 2.4\,MHz. \textbf{c,} The full linewidth ($\kappa/2\pi$) of some resonances of the TM$_{11}$ mode family. The green vertical line in all plots represents the position of the pump laser for the frequency comb generation.}
\end{figure*}

\begin{figure*}[h]
\pdfbookmark[1]{Figure 7}{Fig.7}
\centering
\includegraphics[width=0.7\textwidth]{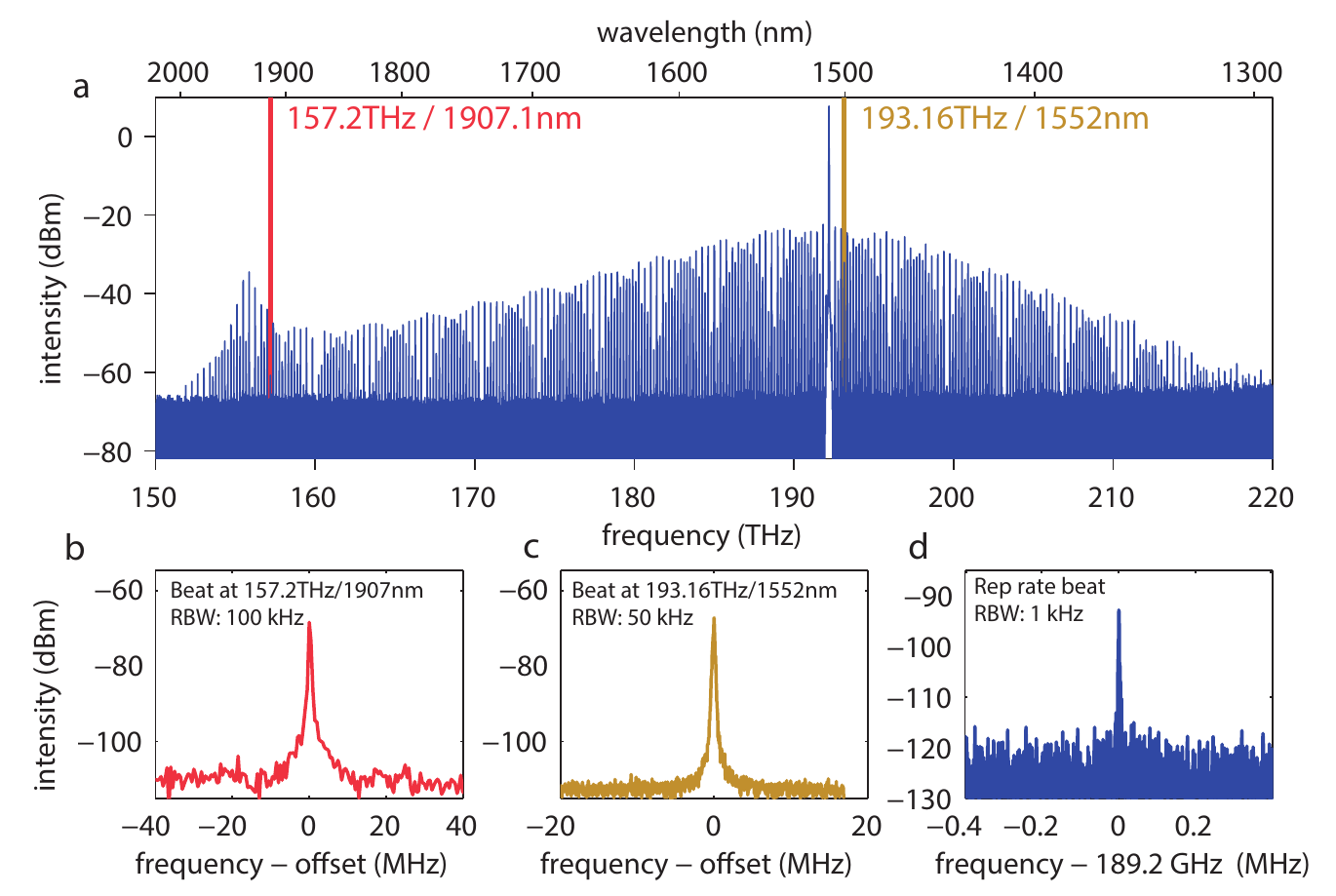}
\caption{\textbf{SI -- Coherence of a two-soliton state in a SiN microresonator.} \textbf{a} Optical spectrum of a two-soliton state measured under very similar conditions and in the same resonance as the data of Fig.\,3. \textbf{b,} The beat note of a narrow linewidth fiber laser at 1907\,nm with the nearest line of the generated frequency comb. \textbf{c,} The beat note of a narrow linewidth fiber laser at 1552\,nm with the nearest line of the generated frequency comb. \textbf{d,} The repetition rate beat note of the generated frequency comb.}
\end{figure*}

\begin{figure*}[h]
\pdfbookmark[1]{Figure 8}{Fig.8}
\centering
\includegraphics[width=0.85\textwidth]{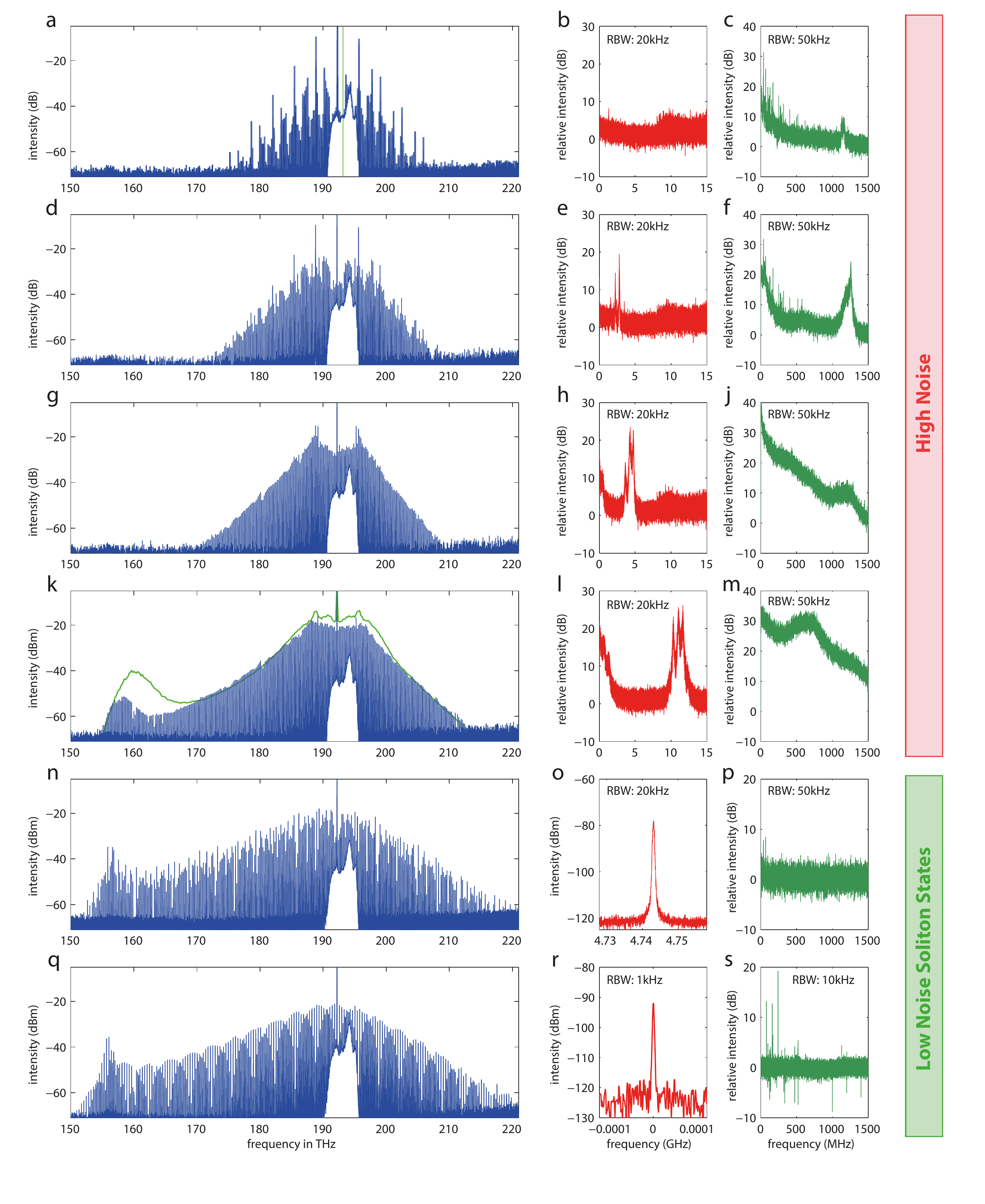}
\caption{{\textbf{SI -- Comparison of a high noise state and solition states in a SiN microresonator.} \textbf{a, d, g, k} The optical spectra of the high noise state for the same resonance as all other spectra and with similar pump power when tuning the pump further into the resonance (from blue to red). The spectrum in \textbf{k} shows a clearly flattened maximum of the soliton induced Cherenkov radiation at around 156\,THz and two maxima to the left and right of the pump at 192.2\,$\pm$\,3\,THz. The green solid line outlines the spectral envelope derived from simulations which agrees well with the experimental data. Reasons of the deviation between simulation and experiment are explained in the main text. \textbf{b, e, h, l} The beat note of a comb line with a narrow linewidth fiber laser at 1552\,nm (green vertical ine in \textbf{a}) in the RF domain (shifting from around 2.5\,GHz in \textbf{e} to around 11\,GHz in \textbf{l}) is very broad and structured. \textbf{c, f, j, m} The amplitude noise of the light after the chip measured up to 1.5\,GHz is clearly elevated. \textbf{n} Optical spectrum of a four soliton state in the same resonance. Although a structure is visible (cf.\ Fig.\,3) the state is low noise. \textbf{o,} The beat note with a narrow linewidth fiber laser at 1552\,nm is well defined and limited by the linewidth of the pump laser. \textbf{p,} The amplitude noise floor of the transmitted light shows no increase over the noise from the pump laser. \textbf{q,} The spectrum for a two-soliton state shows a clear pattern and is also of low noise. \textbf{r,} The repetition rate beat note is narrow and well defined. The data is offset by 189.2\,GHz. \textbf{s,} The amplitude noise in the transmission only shows technical noise at certain frequencies. For all axis labeled ``relative intensity'' the background has been subtracted. The absolute levels of the background are \textbf{m}: --99.6\,dBm, \textbf{p}: --98.5\,dBm, \textbf{s}: --110.8\,dBm.}}
\end{figure*}

\begin{figure*}[h]
\pdfbookmark[1]{Figure 9}{Fig.9}
\centering
\includegraphics[width=0.6\textwidth]{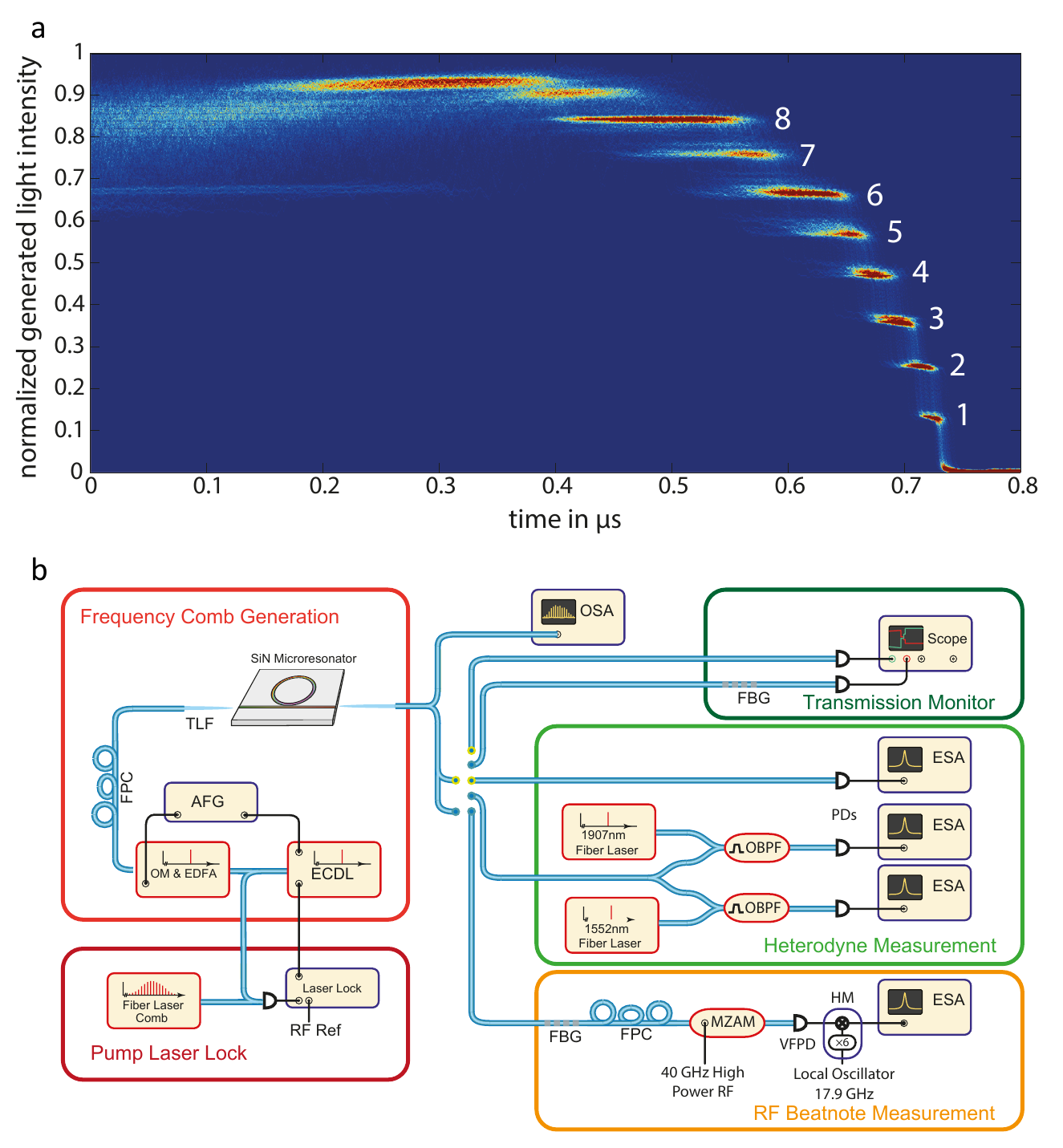}
 \caption{\textbf{SI -- Soliton steps and schematic experimental setup. a,} A color-coded histogram (red denotes high probability, dark blue denotes zero probability) of the recorded steps in the intensity of the converted light (light that is not at the wavelength of the pump) versus laser detuning, revealing steps between different soliton states inside the cavity. {The white numbers indicate the number of solitons for each state.} The transmitted light from the resonator is filtered for the pump and detected, while the laser is scanned through the cavity resonance of the SiN microresonator. \textbf{b,} The setup for microresonator based soliton generation. The microresonator is pumped with CW laser light from an external cavity diode laser (ECDL) that is amplified and modulated in power (bright red box).  The remaining setup is for characterization and stabilization only. AFG, arbitrary function generator; EDFA, erbium-doped fiber amplifier; ESA, electrical spectrum analyzer; FBG, fiber Bragg grating; FPC, fiber polarization controller; HM, harmonic mixer; MZAM, Mach-Zehnder amplitude modulator; OBPF, optical band-pass filter; OM, optical modulators; OSA, optical spectrum analyzer; PD, photodiode; TLF, tapered-lensed fiber; VFPD, very fast photodiode.}
\end{figure*}

\begin{figure*}[h]
\pdfbookmark[1]{Figure 10}{Fig.10}
\centering
\includegraphics[width=0.6\textwidth]{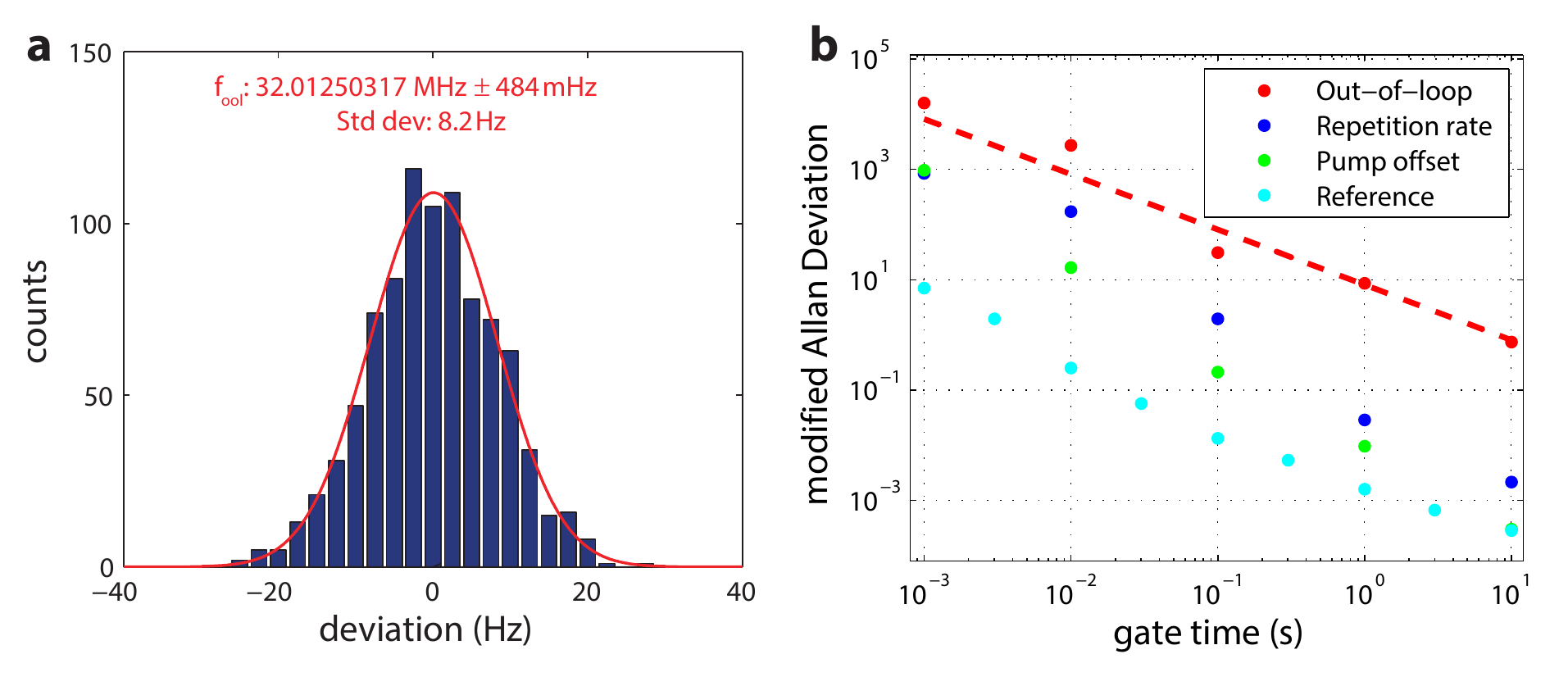}
  \caption{\textbf{SI -- Full stabilization of a single dissipative Kerr soliton in SiN. a} Same data as in Fig.\,4a but for a single soliton state.  Histogram of the counter measurement for the out-of-loop beat of the stabilized microresonator frequency comb with a commercial fiber laser frequency comb. The Gaussian fit gives the exact frequency of the beat. \textbf{b} The modified Allan deviation of the out-of-loop beat as well as the in-loop signals for the two locks of the repetition rate and the pump laser offset of the microresonator frequency comb. All signals average down over the gate time as it should be for coherent signals.}
\end{figure*}

\end{document}